\documentclass[]{aa}
\usepackage{natbib}
\usepackage{graphicx}
\usepackage{txfonts}
\def\HH{\mbox{H$_2$}}

\def\xe{x$_{\rm e}$}         
\def\ne{n$_{\rm e}$}         
\def\nofH{n(H)}         

\def\nH2{n(H$_2$)}

\def\NH2{{\rm N}({\rm H}_2)}
\def\pccc{~{\rm cm}^{-3}} 
 
\def\pcc {~{\rm cm}^{-2}}

\def\Tsub#1 {\mbox{${\rm T}_{\rm #1}$}}
\def\TK  {\Tsub K }
\def\TB  {\Tsub B }

\def\Texc {\Tsub ex }

\def\Tcmb{\Tsub cmb }

 \def\arcmin{\mbox{$^{\prime}$}}
\def\omet{\mbox{$(1-{\rm e}^{-\tau})$}}

\def\p{\mbox{$^+$}}
\def\cotw {\mbox{$^{12}$CO}}
\def\coth {\mbox{$^{13}$CO}}

\def\hcop{\mbox{{HCO\p}}}

\def\WCO{\mbox{W$_{\rm CO}$}}

\def\W13{\mbox{W$_{13}$}}

\def\cch{\mbox{C$_2$H}}

\def\h13cop{\mbox{{H$^{13}$CO\p}}}

\def\c3h2{\mbox{C$_3$H$_2$}}
 
 \def\R0{R$_0$}

\def\ddeg{{}^\circ\kern-.1em}

\def\kms{\mbox{km\,s$^{-1}$}}

\def\bll{BL Lac}

\def\E#1 {$10^{#1}$}
\def\E#1 {E{#1}}
\def\P#1,{$\nH2\TK~=~#1\times~10^4\pccc$~K}
\def\ec#1,#2,#3,{#1\,(#2)\E{#3}}

\def\zoph{$\zeta$ Oph}

\def\H3{\mbox{H$_3$}}

\def\RH2{\mbox{R$_{\rm G}$}}
\def\fH2{\mbox{f$_{\HH}$}}
\def\FH2{\mbox{F$_{\HH}$}}

\def\g13{\mbox{g$_{13}$}}

\newcommand{\supjerome}[1]{}

\title{Rotational excitation of simple polar molecules by \HH\ and electrons in diffuse clouds}
\author{H. S. Liszt\inst{1}}
\institute{National Radio Astronomy Observatory,
           520 Edgemont Road,
           Charlottesville, VA,
           USA 22903-2475}

\begin{document}
\date{received \today}
\offprints{H. S. Liszt}
\mail{hliszt@nrao.edu}
%
% \abstract{}{}{}{}{} 
% 5 {} token are mandatory
\abstract
  % context heading (optional) leave it empty if necessary  
 {Emission from strongly-polar molecules could be a probe
  of physical conditions in diffuse molecular gas.}
  % aims heading (mandatory)
{We wish to provide basic information needed to interpret emission
 from molecules having higher dipole moments than CO, originating in diffuse clouds
 where the density is relatively low and the temperature and electron fraction are 
 relatively high compared to dark clouds.}
  % methods heading (mandatory)
 {Parameter studies in LVG models are used to show how the
 low-lying rotational transitions of common polar
 molecules \hcop, HCN and CS vary with number density, column density 
 and electron fraction; with molecular properties such as the charge state 
 and permanent dipole moment; and with observational details such as the 
  transition that is observed.  Physically-based models are used to check 
 the parameter studies and provide a basis for relating the few extant
 observations.}
  %results heading (mandatory)
{ Parameter studies of LVG radiative transfer models show that 
 lines of polar molecules are uniformly brighter for ions, for lower 
  J-values and for higher dipole moments.  Excitation by
  electrons is more important for J=1-0 lines and contributes
  rather less to the brightness of CS J=2-1 lines.  If abundances 
  are like those seen in absorption, the \hcop\ J=1-0 line will be the 
  brightest line after CO, followed by HCN (1-0) and CS (2-1).  
 Because of the very weak rotational excitation in diffuse clouds,
  emission brightnesses and molecular column densities retain a nearly-linear
 proportionality under fixed physical conditions,  even when transitions are 
  quite optically thick; this implies that changes in relative intensities among 
 different species can be used to infer changes in their relative abundances.} 
 %conclusions headiing (optional), leave it empty if necessary
{}

%intensity proportional to colum density even when optically thick
%impossibility of strong electron excitation because x_e falls as n(h)
%increases
%
\keywords{ interstellar medium -- molecules }

\authorrunning{H. S. Liszt}
\titlerunning{The importance of being polar}

\maketitle

%s1
\section{Introduction}

The rotational excitation of CN in diffuse clouds was discussed as soon as 
CN was identified in the diffuse ISM by \cite{McK40}.  He calculated the
``effective'' or ``rotational''  temperature of interstellar space to 
be no more than 2.7 K and asked ``if, indeed, the concept of such a 
temperature in a region with so low a density of both matter and 
radiation has any meaning.''  The efficacy of CN as an early 
probe of the cosmic microwave background  temperature was confirmed 
by its absence in several searches for mm-wave emission 
\citep{Tha72,CraHeg+89,BlaVan91}, but gaining insight into the 
structure of the absorbing gas requires more than upper limits 
on the excitation temperature of such a trace molecule.  Moreover,
actually detecting emission from various species among the 
many now known to exist in diffuse clouds may even permit 
imaging of the structure and chemistry of the host gas.

CO emission from diffuse clouds has long been known (see \cite{KnaJur76}
although several of the lines discussed there are telluric as noted
in \cite{Lis09COSurvey}) and measurement of 
its excitation temperatures in absorption provides direct 
information on the ambient partial thermal pressure
p/k $\approx$  n(\HH)\TK\ $= 1-10\times10^{3}\pccc $ K 
\citep{SmiSte+78,LisLuc98}.  However, the mm-wave brightness of CO emission
from diffuse gas is for the most part proportional only to the total 
CO column density N(CO) \citep{GolKwa74,Lis07CO}, somewhat leaving
the determination of the physical properties of the medium to other
means, which for the most part remain to be explored.

Emission from diffuse clouds is accessible in 18cm OH emission \citep{Cru79} 
and 9cm CH emission \citep{MagOne+98,MagCha+03}, but the lines are weak, 
the spatial resolution is relatively poor and the emission arises from states 
that are difficult to model or interpret \citep{FelRou96,LisLuc96,LisLuc02} .  
Determination of the internal properties 
of diffuse gas would be well-served by wider detection of mm-wave emission of 
molecules more strongly polar than CO, whose emission brightness might
depend more directly on the number density and electron fraction in
addition to the molecular column density,  and which could be observed with 
arcminute or better spatial resolution.  Moreover, receivers have recently
undergone drastic improvement in the 3mm band, so that the weakness of
lines (see below) should not be such a barrier to future progress.

Emission from strongly polar molecules has been seen in limited fashion
in diffuse gas, in \hcop\ \citep{LisLuc94,LucLis96,Lis97,KopGer+96,FalPin+06}, 
CS \citep{DrdKna+89} and HCN \citep{Lis97}.  Where the \hcop\ column density 
and optical depth are known in absorption, the weak \hcop\ emission from 
opaque lines requires a small excitation temperature given by the line 
brightness, so that the density can be discussed independent of the column 
density \citep{LucLis96,LisLuc00}.  
More generally, the density of colliding partners and the column density of
emitting molecules are not separable from observations of a single
transition, but surveys of the relative abundances of the various molecules 
in absorption might allow some of this unfortunate degeneracy to be relieved.
Observing the relative intensities of various species could a means of deriving
the ambient physical conditions if the relative abundances of those
species are sufficiently well constrained by other means, and vice versa.

The plan of this work is as follows.  In Sect. 2 we lay out the basic 
elements of model calculations of rotational excitation by collisions 
with electrons and \HH\ in diffuse gas for species with different and 
varying permanent dipole moment and abundance.  The first of these studies, 
described in Sect. 3, is a purely parametric study using LVG radiative 
transfer \citep{GolKwa74} to limn the borders of the relevant parameter 
space.  The second is a more physically-based study of excitation in 
models where the thermal and chemical balance and CO brightness are 
calculated in some detail and the results are relevant to interpretation 
of emission line observations of polar species when the only other 
or {\it a priori} information available on the host cloud is the strength 
of CO emission.  The results of these studies are presented along with a 
summary of the extant observations in Sect. 4.  The results of this
work are summarized in Sect. 5.

%s2
\section{Models of excitation of polar molecules in diffuse clouds}

Molecule-bearing diffuse clouds are marked by elevated temperatures:
$<$\TK$>$ = 70-80 K for \HH\ \citep{SavDra+77,RacSno+02} generally, 
while somewhat lower temperatures are found for regions with higher 
N(CO) = $10^{14} - 10^{16} \pcc$ \citep{SonWel+07,SheRog+07}.  
 They are also marked by 
modest density n(H) $ = 30 - 500 \pccc$ and high electron fraction 
n(e)/n(H) $\approx 1-2\times 10^{-4}$, the latter corresponding to 
the free gas phase abundance of elemental carbon \citep{SofLau+04} 
and a contribution from cosmic-ray ionization of hydrogen \citep{Lis03}.
Excitation calculations also must take into account that the J=1-0 rotational 
transition of a species having a permanent dipole moment $\mu$ Debye
(1D = $10^{-18}$ esu) will have $\tau = 1$ at column density 
per unit velocity dN/dV $\approx 2\times 10^{13} \mu^{-2} \pcc$ (\kms)$^{-1}$,
as is observed for many species.

When the excitation is as weak as that which occurs in diffuse clouds for
strongly-polar molecules, or said somewhat differently, so very
sub-thermal (even in CO although it is much more strongly excited), emission
line brightnesses are linearly proportional to column density even when
the line optical depth is large.   This occurs because, with so little 
collisional excitation, there is also little collisional de-excitation 
and most collisional excitations are followed by radiative de-excitation
(Lyman-$\alpha$ radiative transfer is an extreme example of the same phenomenon).  
Photons escape the medium even if they are scattered and re-absorbed many 
times \citep{GolKwa74}.  For CO the brightness is also insensitive to the 
ambient density at fixed N(CO) \citep{GolKwa74,Lis07CO} and the electron fraction
is unimportant because of the small dipole moment (0.11D).

\subsection{Excitation by electrons}

Excitation of strongly polar molecules by electrons has a strongly 
dipole character ($\Delta$J$ = \pm 1$) and the rates for such transitions 
are quite accurately represented in closed form for molecular ions 
\citep{DicFlo81,BhaBha+81} and neutrals \citep{DicPhi+77} having dipole moments 
$\mu \ga 0.5$D. The two formulations of the e-ion excitation rates are entirely 
equivalent although somewhat differently expressed (see also \cite{NeuDal89}).  
More accurate rates for e-\hcop\ and e-HCN collisions
including $|\Delta{\rm J}| <> 1$ transitions have recently been 
calculated by \cite{FauTen+07} and \cite{FauVar+07}, respectively,  and 
for CS by \cite{VarFau+10} and a comparison between calculations using them 
and the older  rates in closed form is given below for \hcop\ in Fig. 3.  

\subsection{Excitation of strongly-polar molecules by neutral particles}

Excitation rates for \HH-\hcop\ interactions were extensively recalculated
by \cite{Flo99}.  Other species explictly discussed here are CS \citep{GreCha78}
and HCN \citep{GreTha74,MonStu86}.  The dependence of these collisional
rate constants
on the dipole moment of the target molecule is ignored here because they
should be relatively weak compared to the other dependencies such as the 
electron-excitation rates and the spontaneous emission rates.

\subsection{Excitation of carbon monoxide}

The excitation of CO in diffuse clouds was recently discussed by \cite{Lis06}
and \cite{Lis07CO}.  Note that the very large H-CO excitation rates
calculated by \cite{BalYan+02} and extensively discussed by \cite{Lis06} were
subsequently repudiated \citep{SheYan+07}.  Because of its low dipole moment
CO is largely unaffected by the electron fraction and this is illustrated in
some of the parameter studies shown below where the molecular dipole moment
is artificially varied to take advantage of the existence of analytic formulae
for the electron excitation rates.  For strongly sub-thermal excitation,
the excitation temperature of the J=1-0 transition of CO is proportional to
the ambient partial thermal pressure of \HH, p/k = \nH2\TK, not to either \TK\ or
\nH2\ individually.

\subsection{Permanent dipole moments}

The permanent dipole moments used here are: CO, 0.112D; \hcop, 3.889 D; CS, 1.96 D;
and HCN, 2.98 D.  In our earlier work we used a value $\mu$ = 4.07 D for \hcop, 
which gave smaller \hcop\ column densities by a factor (4.07/3.889)$^2$ = 1.10
for a given integrated optical depth.

%1
\begin{figure*}
\includegraphics[height=13cm]{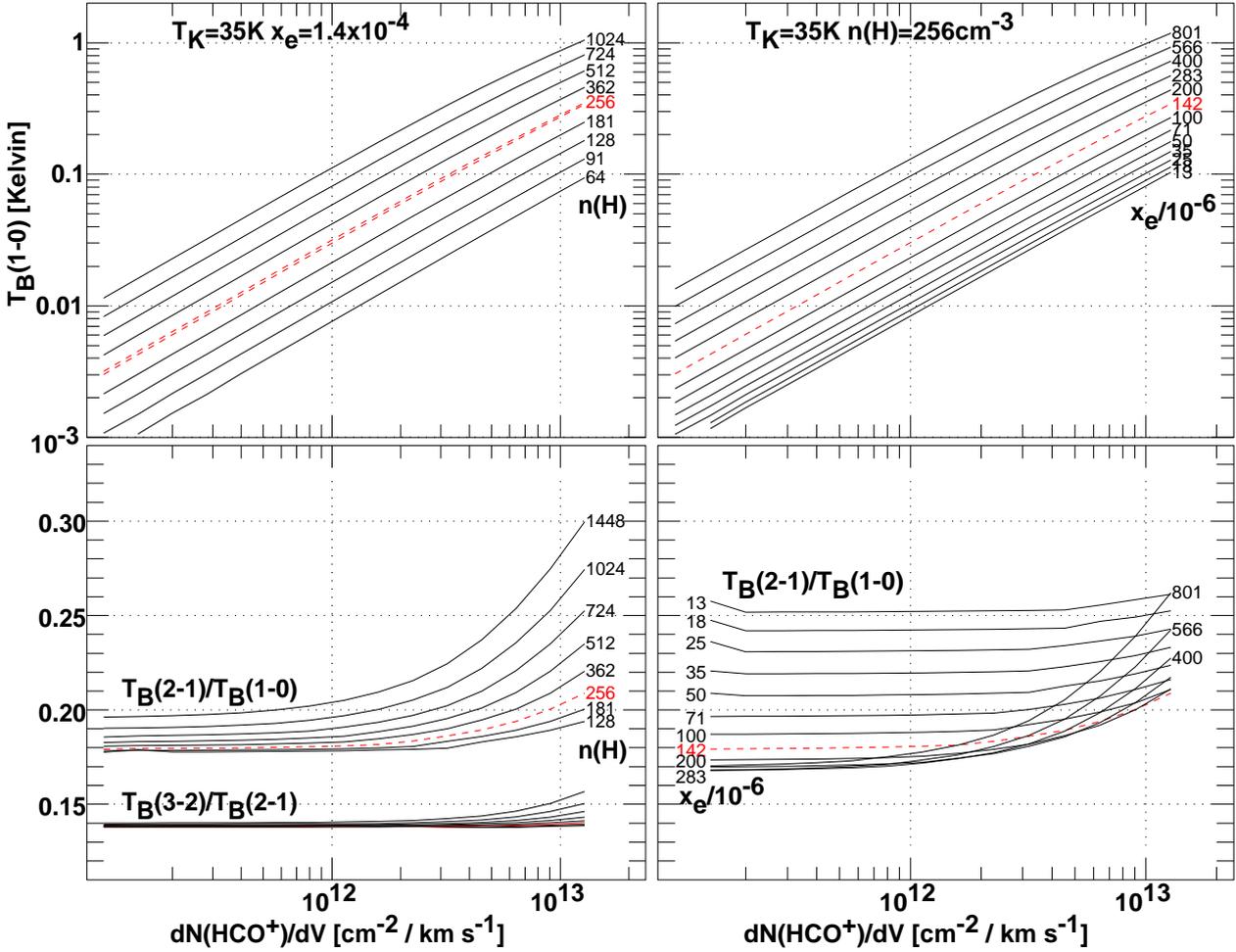}
\caption[]{Rotational excitation of \hcop\ by \HH\ and electrons in 
diffuse gas at moderate density in the LVG approximation.   The horizontal 
axis of all panels is the column density per unit velocity of \hcop.  The 
top panels show the brightness temperature above the cosmic microwave
background of the J=1-0 line , at left for fixed electron fraction 
\xe\ = \ne/\nofH\ = $1.4\times10^{-4}$ and varying hydrogen number 
density \nofH\ = 2\nH2, at right for
varying electron fraction at \nofH\ = $256 \pcc$.  The lower panels show
 line brightness ratios in the same manner.  At upper left the calculation
for \nofH\  $= 256 \pccc$ is shown twice, for \TK\ = 15 (which lies below) and 
35 K. Note that the lower panels in each column have a linear ordinate scale.}
\end{figure*}

\subsection{Radiative transfer}

 Strongly polar molecules typically become optically thick in their low-lying 
transitions for column densities around dN/dV $ \ga 10^{12}\pcc$ (\kms)$^{-1}$, 
which is typical of  actual observed abundances.  Therefore it is necessary to 
treat the radiative transfer in some fashion. Sect. 3 presents a series of 
parameter studies in the usual LVG approximation \citep{GolKwa74} which amounts 
to solving the full, coupled, excitation rate equations in the escape probability 
approximation,
taking  q = \omet$/\tau$ as the photon escape probability for a transition 
whose optical depth is $\tau$ and setting the line source function
S$_\nu$ = q B$_\nu$(\Tcmb) + (1-q) B$_\nu$(\Texc).  In these expressions
\Texc\ is the transition excitation temperature and  B$_\nu$ is
the Planck function.  As noted by \cite{GolKwa74}, using the escape probability 
formulation transforms the rate equations
so that they resemble the optically thin limit but with all the radiative rates 
(basically, the Einstein A-values) multiplied by q.

Section 4 presents a more detailed calculation for small uniform density gas spheres 
in which the radiative transfer is calculated numerically in the microturbulent 
approximation.

\subsection{Observed relative abundances}

The abundances of \hcop, HCN and CS observed in absorption at mm-wavelengths are
summarized in Fig. A.1 of Appendix A.  The calculations here were done with
X(CS) = X(HCN) = X(\hcop) = $ 2\times10^{-9}$.

\subsection{Molecules not considered here}

The discussion here is limited to a few of the simplest species that
are detected in diffuse gas. Failed searches for emission from CN 
($\mu=1.45$D) were noted in the Introduction but its intensity is severely 
diluted by hyperfine structure; HCN, whose abundance with respect to CN is fixed 
in diffuse gas  \citep{LisLuc01}, has a twice-larger dipole moment and is brighter 
in emission, leaving little reason to continue the old quest for CN. \cch\ is 
ubiquitous in diffuse gas  \citep{LucLis00C2H,GerKaz+11}, although its emission 
brightness will be diluted by hyperfine structure. The dipole moment of \cch\ 
is 0.78 D. 

\section{Parameter studies}

\subsection{The \hcop\ J=1-0 transition}

Figure 1 shows a composite of results for a calculation of \hcop\ excitation, where
the horizonal axis is the column density per unit velocity. As context, note that the
\hcop\ J=1-0 line has optical depth $\tau = 1$ for dN/dV $ = 1.1 \times 10^{12}
\pcc$ (\kms)$^{-1}$ and many other molecules like HCN \citep{LisLuc01} and CS \citep{LucLis02}
increase in abundance relative to \hcop\ for N(\hcop) $ \approx 1-2 \times 10^{12} \pcc$.
The largest values we have seen in absorption for individual kinematic components
is N(\hcop) $ \approx 4-7\times 10^{12} \pcc$ \citep{LucLis96} and typical line profile
FWHM are 1-1.5 \kms.  A typical fractional abundance for \hcop\ is X(\hcop) =
n(\hcop)/N(\HH) = $2-3\times 10^{-9}$ \citep{LisPet+10}. 

At the top in Fig. 1 the brightness of the J=1-0 line is plotted for a wide range
of hydrogen number density at fixed \xe\ = \ne/n$_{\rm H} = 1.4 \times 10^{-4}$ and at 
right for varying \xe\ at \nofH\ = 256 $\pccc$.  Note that  \nofH\ = 2 \nH2\ in the LVG 
calculations of Fig. 1-3 and the use of \nofH\ rather than \nH2\ is for the sake of 
consistency with the more detailed calculations (see Sect. 4) in which \nofH\ is 
stipulated and the H-\HH\ equilibrium is actually calculated.  Clearly the 
line brightnesses are linearly proportional to the column density for both 
optically thick and thin emission as for sub-thermally excited
 CO (see Fig. 6 of \cite{Lis07CO}), but they are also
sensitive to the density of \HH\ and electrons.  At left, the calculation
for n(H) $= 256 \pccc$ has actually been repeated for \TK\ = 35 and 15 K to
illustrate the weak temperature dependence of electron excitation of the J=1-0
transition.  This simple calculation accounts for the weak observed \hcop\ 
lines seen near absorption line background targets \citep{LucLis96} 
with typical values of the electron fraction at modest n(H) $\la 300 \pccc$.  
Note that $\tau \approx 1$ near the horizontal mid-point of each panel,
so that $\tau > 10$ at the rightmost edge.  Some deviation from linearity
is seen at the highest densities and electron fractions where the excitation
is stronger and more molecules are supported in higher levels of the rotation
ladder.

At top right in Fig. 1 it is seen that electron excitation accounts for 
some 75\% of the 
line brightness under these nominal conditions.  Electron fractions much 
higher than 10$^{-4}$ are hard to maintain in denser gas given the free carbon 
abundance $1.4 \times 10^{-4}$ and the electron fraction in physically-based 
models actually declines noticeably for \nofH\ $> 100 \pccc$, although carbon 
remains almost fully-ionized (see Sect. 4).

In the lower panels we show the brightness of the higher lying lines 
(the J=2-1 line is often inaccessible near z = 0 owing to atmospheric
absorption).  At moderate column density they are somewhat more dependent on 
the electron fraction than the number density  (see at right) while the opposite 
is true at high density.  At present, detecting emission from any but 
the J=1-0 line of \hcop\ seems unlikely for diffuse clouds.

\subsection{CS J=2-1}

Figure 2 shows the brightnesses of the lowest three lines of CS, 
calculated using the closed form of the electron-neutral molecule 
excitation rates (also see \cite{DrdKna+89}).  As for \hcop\ the 
CS lines decline strongly in brightness with each successive increase 
in J and although the \hcop\ and CS J=1-0 lines will be about equally
bright at the same column density, the predicted CS J=2-1 lines are about
a factor 6 weaker.  About  equal column densities are typical for N(\hcop) $\ga 10^{12}\pcc$
(see Fig. A.1).  Weakness of the CS emission is due to a variety of 
factors, CS being neutral and less strongly polar (1.96 vs. 3.89 D), but 
mostly the difference arises from comparison of J=2-1 and J=1-0 transitions.  
Less of the 
line brightness is due to electrons in the CS J=2-1 transition than
for \hcop\ J=1-0, perhaps half (top right), and very little at all 
for the CS J=3-2 line given the difficulty of moving population into 
higher J-levels with only $\Delta$J $=\pm1$ transitions.  As shown in Fig. 
2, the brightnesses of the higher-lying CS transitions are more sensitive 
to temperature  because more of the excitation is due to \HH\ and 
the rotation levels lie higher in energy so that purely thermodynamic 
considerations become more important.

%2
\begin{figure*}
\includegraphics[height=13cm]{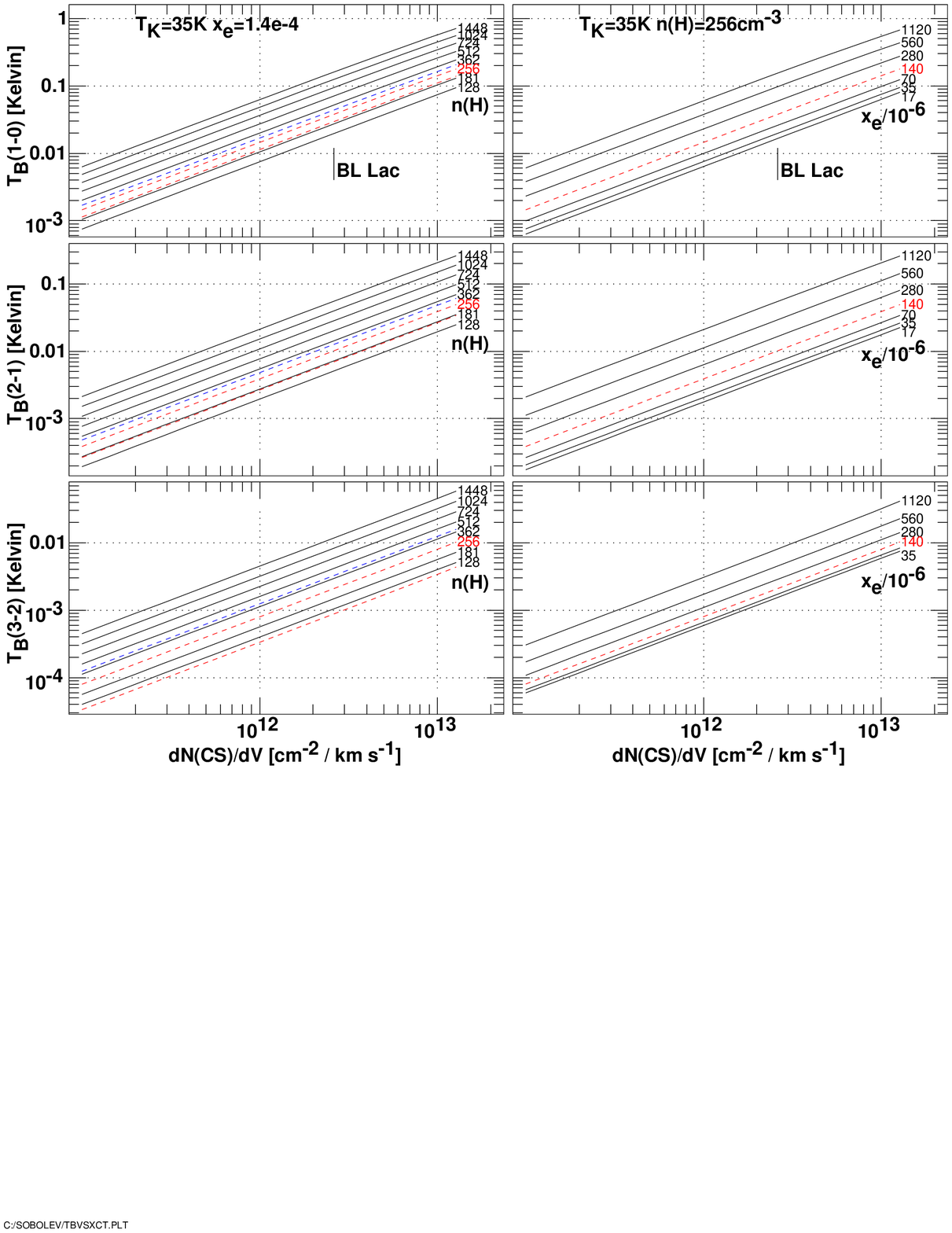}
\caption[]{Rotational excitation of CS by \HH\ and electrons in diffuse
gas at moderate density, much as in Fig. 1.  From top down
the panels show the brightness temperature of the CS J=1-0, J=2-1
and J=3-2 transitions.  At left the calculation for \nofH\ 
$= 256 \pccc$ is shown for 15, 35 and 55 K in all panels.  The observed
value toward \bll\ is indicated.}
\end{figure*}

\subsection{The importance of being polar}

To illustrate the influences at work in these excitation 
studies owing to the structure of the individual species,
we performed another sort of parameter study which takes advantage
of the availability of closed forms for the electron excitation
rates.  The excitation and line brightness are calculated for 
various transitions and
molecules, but the permanent dipole moment is treated as a variable
whereever it appears explicitly.  In particular we consider the 
J=2-1 line of CS  and the 1-0 transitions of  HCN and \hcop.

Figure 3 shows  the excitation temperatures (at left) and line 
brightness (at right) as the permanent dipole moment varies from
0.1 to 8 D, for the default values employed in Fig. 1-2 (n$_{\rm H}$
= 256 $\pccc$, \ne/n$_{\rm H}$ = $1.4\times 10^{-4}$, \TK = 35 K),
for hypothetical species whose energy level spacings {\it etc.} are
the same as for \hcop, HCN and CS.  Highlighted are the series
of models at those values closest (actually very very close) to the 
true dipole moments.  
For \hcop, results are also shown for the correct permanent dipole
moment and the more accurate recent excitation rates of \cite{FauTen+07}.
The \hcop\ lines are somewhat brighter and more highly excited when the
more detailed electron excitation scheme is employed; larger
differences would appear for the higher-lying lines.

The behaviour seen in Fig. 3 is curious in some regards.  Excitation 
temperatures decline monotonically with $\mu$ for optically thin lines
and then increase when both the dipole moment and optical depth grow large but
none of this behavior is apparent in the line brightness:  in 
all cases the lines are simply brighter for higher dipole moments.
It is generally noted that electron excitation provides about the
same degree of excitation for all species independent of $\mu$
because the excitation rates and spontaneous emission 
rates both vary as $\mu^2$. However the current cases are not 
pure, more so for CS J=2-1, and electron excitation generally 
dominates only when the dipole moments are large even for the 
species which actually are more strongly polar.  The decreasing
excitation temperatures at small $\mu$ occur because the excitation 
is dominated by neutrals so that increasing $\mu$ mostly serves to
cause higher but uncompensated spontaneous decay rates.  Eventually,
increasing the dipole moment raises the line optical depths to
the point that radiative pumping props up the excitation temperatures.

%3
\begin{figure*}
\includegraphics[height=16cm]{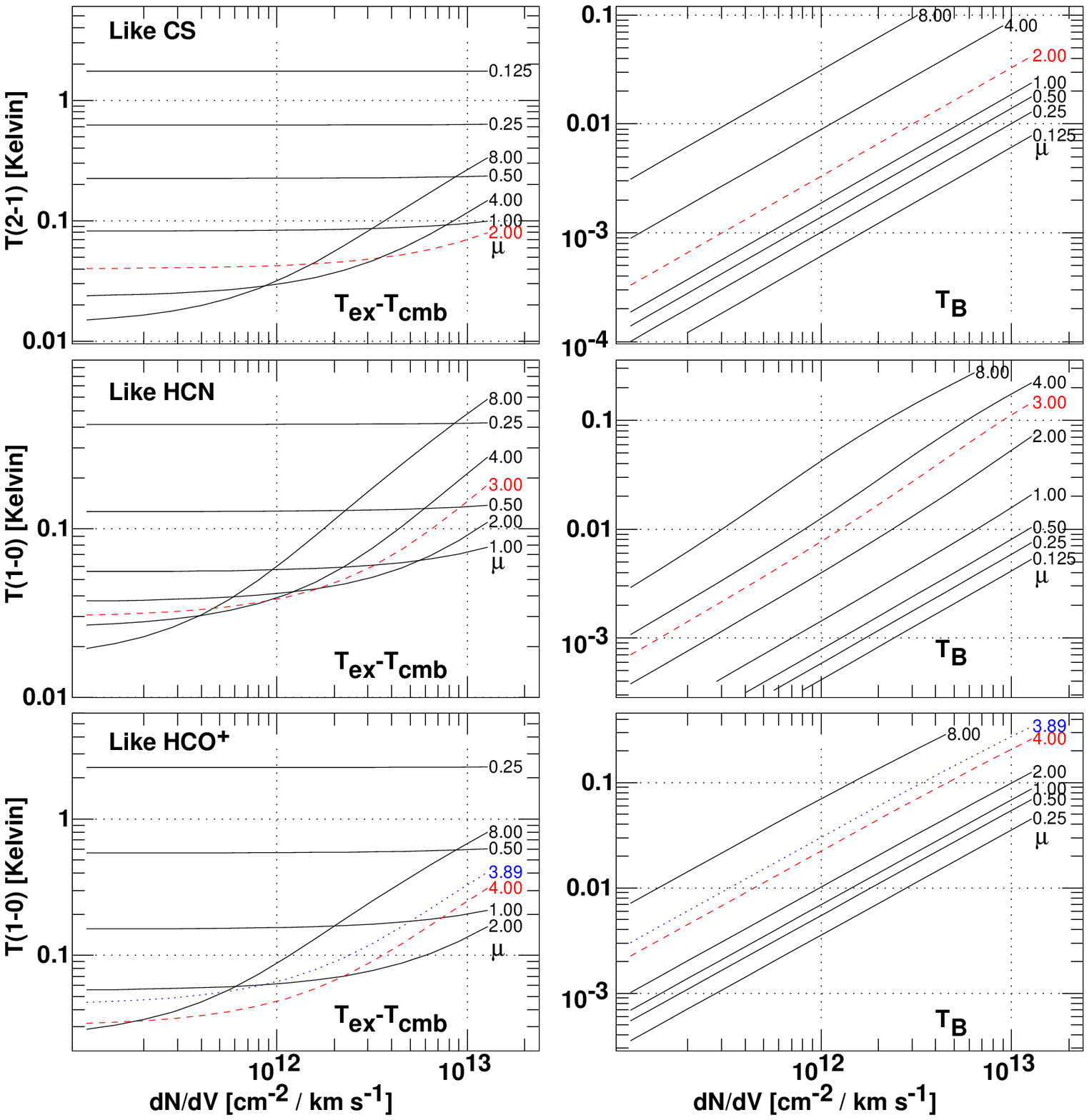}
\caption[]{Dependence of the rotational excitation upon permanent
dipole moment $\mu$ (Debye) at \nofH\ = 2\nH2 = 256 $\pccc$, \TK = 35 K and
\xe\ = $1.4 \times 10^{-4}$ as before. The horizontal axis is column density 
per unit velocity and vertical axes are line excitation temperatures (at left) 
and  brightness temperatures above the cosmic background.  The 
CS-like species at top has the same rotational energy spacing
and \HH-excitation rates as CS, but the dipole moment is varied
wherever it appears analytically in expressions for the line optical
depth, brightness {\it etc.} and electron-neutral molecule excitation 
rates.  Panels in the middle row are for the strongest hyperfine component
of the J=1-0 line of a neutral HCN analog, and the lowest two panels
in each column pertain to the J=1-0 line of a molecular ion analog of \hcop\
(compare with Fig. 1).
In each row one series of calculations is shown by red dotted lines;
by coincidence those use values very close to the actual dipole moments 
of CS (1.96 D), HCN (2.98 D) and \hcop\ (3.89D). The dotted blue curve 
in each of the lowest two panels uses the correct dipole moment and the 
recent electron excitation rates of \cite{FauTen+07}}.
\end{figure*}

\section{Existing observations of \hcop, HCN and CS emission}

To compare observations and datasets on something like a common scale,
we searched the literature for observations of polar species with 
accompanying \cotw\ measurements and these are shown in Fig. 4.
In the following we denote the integrated brightness of the
J=1-0 line by \WCO, while those CS and HCN are W(CS) and W(HCN). 

\subsection{\hcop}
 
Shown in Fig. 4 are observations of \hcop\ and \cotw\ J=1-0 from 
the \hcop\ absorption line survey of \cite{LucLis96} (for CO also see
\cite{LisLuc98}) and emission line observations of the clouds occulting 
\zoph\ by \cite{KopGer+96} and \cite{Lis97}.
\cite{FalPin+06} only reported \coth\ measurements to accompany their
observations of \hcop\ in Polaris. The 
general run of \hcop\ brightnesses in the range W(\hcop) = 0.01 - 0.1 K \kms\ is
very much in line with the model results in Fig 1 for n(H) $= 256 \pccc$
and N(\hcop) $= 0.3 - 3 \times 10^{12}$ as typically observed.
\hcop\ appears in diffuse gas at a level W(\hcop) $\simeq$ \WCO/100.

\subsection{HCN}

HCN emission was probably detected at the position of peak \hcop\ emission
South of \zoph\ \citep{Lis97} at about 1/3 the strength of \hcop; this
datapoint appears in Fig. 4 at \WCO\ = 6 K \kms.  It is in line with 
expectations for approximately equal column densities of \hcop\ and HCN
as observed in absorption, see Fig. A.1.

\subsection{CS}

Observations of CS J=2-1 were made by \cite{Lis97} at positions
around the sightline to \zoph.  Where \hcop\ emission was strongest 
30\arcmin\ South of the star, with W(\hcop) = 0.09 K \kms and W(CO) = 
6 K \kms, CS was undetected at a ($3\sigma$) level 5 times weaker.  
CS was also undetected toward the star, in line with expectations 
for the calculated excitation conditions in Figs. 1 and 2 with 
N(CS) $\approx$ N(\hcop) as observed in absorption.  More recently 
(to be published), we failed to detect CS emission at several other 
locations in the Southern portions of the \zoph\ cloud, including 
in directions showing 15 K CO lines. 

However, much 
stronger CS emission was reported by \cite{DrdKna+89} along 
several sightlines toward bright stars used as targets for optical absorption 
line studies.  It is 
possible that the strong CS emission arises from material located behind 
the target stars, as is often the case for CO emission in such surveys 
\citep{Lis09COSurvey} but the accompanying \hcop\ lines would be 
very bright indeed, and anomalously bright with respect to CO, if they 
were several times stronger than the observed CS.  The N(CS)/N(\hcop) ratio 
may vary over a wider range than was apparent toward the set of
background sources we surveyed in absorption, 
where N(CS)/N(\hcop) $<3$ according to Fig. A.1.

\subsection{Physical models}

In practice, emission lines of CO may well be all that is available
to infer cloud properties when \hcop\ or other polar species are
sought and that is the common meaure we applied to compare observations
in Fig. 4.  To bridge the gap between observation and the LVG parameter
studies described in Sect. 3, we extended our earlier models of \HH\
and CO formation \citep{Lis07CO} to calculate the emergent line brightness
of \hcop\ and CS as well as CO.  The models self-consistently determine
the thermal equilibrium \citep{WolHol+95} and equilibrium abundances of
\HH\ and CO as a function of radius in small spherical gas clots of given
number and column density n(H) and N(H), under the 
assumption\footnote{This general methodology is known as ``faith-based chemistry''} that CO
forms from the ordinary thermal recombination of a fixed relative
abundance of \hcop, X(\hcop) $=2\times 10^{-9}$.  The emergent
line brightness is calculated for a microturbulent medium of
fixed linewidth.  For instance, Fig. 6 of \cite{Lis07CO}
shows how \WCO\ $\propto$ N(CO) for 0.1 K \kms\ $\la$ \WCO\ $\la$
10 K \kms,  $10^{14} \pcc \la$ N(CO) $\la 10^{16} \pcc$ (as observed)
even though $\tau \simeq 1$ at N(CO) $\simeq 10^{15}\pcc$.

Shown in Fig. 4 are two series of models for \hcop\ and for CS. Each 
curve is for a fixed central column density N(H) = 0.7 or 2 
$\times 10^{21}\pcc$ and the internal number density of hydrogen 
increases to the right in marked steps of 2$^{0.5}$ beginning with 
n(H) = 128 $\pccc$ at the left-most point.  The relative abundances 
of CS and \hcop\ were taken equal, $2\times 10^{-9}$.  The \hcop\ and CS 
brightnesses are slightly lower at higher N(H) and fixed \WCO\ because 
models with higher N(H) form CO more readily and have slightly lower 
electron fractions.  Models with higher N(H) have brighter CS and 
\hcop\ lines at the same number density (i.e. at their respective 
left-hand sides) but only because they have commensurately higher 
CS and \hcop\ column densities.

The models reproduce the general run of observations of \hcop\ but 
the calculated W(\hcop)/\WCO\ ratios are basically fixed by the \hcop\ 
recombination chemistry and the empirically-determined \hcop\ 
abundance and are very difficult to fine-tune.  The W(\hcop)/\WCO\ 
brightness ratio actually decreases slightly when X(\hcop) increases
because the CO abundance increases even faster.  This is opposite 
to the linear, proportional variation of W(CS) with X(CS) that
occurs because changing X(CS) does not affect \WCO.

The physically-based calculation confirms the insight from the LVG
parameter studies that J=2-1 lines of CS will be much dimmer than J=1-0 
\hcop\ when X(CS) = X(\hcop) but the two model curves for CS 
are almost coincident with each other in this parametrization (as
is also nearly the case for \hcop).  This means  
that the only way of
effecting a strong change in the CS/\hcop\ brightness ratio (at
fixed \WCO) will be to vary X(CS).  Although CS and \hcop\ were not
measured along the same directions in Fig. 4 with strong CS emission, the
existence of comparably bright \hcop\ and CS lines at similar values of 
\WCO\ strongly implies that X(CS) $>>$ X(\hcop).

%4
\begin{figure}
\includegraphics[height=8.2cm]{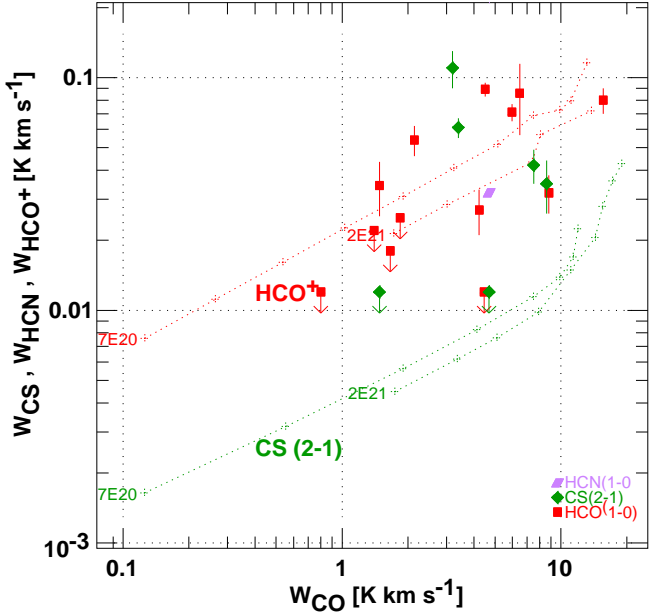}
\caption[]{Integrated emission line brightnesses for
\hcop\ J=1-0 (red boxes), CS J=2-1 (green diamonds) and HCN J=1-0 
(magenta parallelogram) observed in diffuse gas, plotted against
the integrated brightness of CO J=1-0.   Data sources are 
given in Sect. 4.1-4.3.  Chained red and green curves are 
model results as discussed in Sect. 4.4.  For both \hcop\ (red) and
CS (green), calculations are shown for clouds with N(H) =
0.7 and 2.0 $\times 10^{21} ~\pcc$ with  number density n(H) increasing
from  128 $\pccc$ at left in marked steps of 2$^{1/2}$. 
X(\hcop) = X(CS) $= 2\times 10^{-9}$ was assumed; the CS brightness 
increases linearly with increased abundance X(CS) at fixed \WCO\ while
that of \hcop\ declines very slightly, as discussed in 
Sect 4.4 of the text.}
\end{figure}

\section{Summary}

In Sect. 3 (see Figs. 1-3) we discussed a series of parameter studies based on LVG
radiative transfer calculations for actual molecules (\hcop, HCN, CS) 
and a series of synthetic analogues with varying permanent dipole moments. 
In diffuse molecular gas at weak excitation, lines of polar molecules 
are uniformly brighter for ions, for lower J-values and for higher dipole 
moments.  Excitation by electrons is more important for ions and for 
J=1-0 lines and contributes rather less to the brightness of J=2-1 
lines. The fact that \hcop\ is observed in its J=1-0 line 
and at about the same column density as HCN or CS should  make it
several times brighter in emission than HCN J=1-0 or CS J=2-1.
Low levels of \hcop\ J=1-0 emission (\TB\ $ \approx 0.01 - 0.03$ K) 
should be ubiqitous in diffuse gas if (as observed) \hcop\ is itself 
ubiquitous at column densities around N(\hcop) = $10^{12}\pcc$.   Of course
abundances play a role in whether a species is detectable and
the relative abundances of \hcop\ and CS or HCN differ (the CS/HCN
ratio varies less, see Fig. A.1) and may vary by amounts which are 
also comparable to the differences in brightness induced by the 
other factors discussed here.

Perhaps the most important point to take away from the simulations 
is the linear proportionality between intensity and 
column density that applies to all species and low-lying transitions when 
the excitation is as weak as it is in diffuse gas.  This is true almost 
without regard for optical depth over the range of column densities that 
are typical of observed diffuse 
clouds\footnote{This also applies to CO, although the discussion here is
really concerned with more strongly polar species.} .
This proportionality means that model line brightnesses calculated
under given physical conditions can simply be scaled over a very wide range 
of molecular column density in many cases (a notable exception is when
the brightnesses of \hcop\ and CO are compared, see Sect. 4). But it 
also implies that when different species are observed along the same 
lines of sight (more precisely, under the same physical conditions), 
their relative brightnesses are set by their relative abundances. 
Variations in their relative brightnesses from position to position 
result from variations in their relative abundances, because variations
in physical conditions do not shift the pattern of relative brightnesses
appreciably (compare Fig. 1 and 2).  Observing, say, 
CS J=2-1 and \hcop\ J=1-0 lines of comparable brightness along the 
same line of sight really implies that N(CS)/N(\hcop) $>>1$, a situation
that has never been observed in our absorption line work.

We gathered and compared existing observations of mm-wave emission
from polar species in Sect. 4.  To put the different datasets 
on a command basis we plotted their brightneess against \WCO,
the integrated brightness of the J=1-0 CO line.  \hcop\ emission at
a level 1-2\% of \WCO\ is typical.  In line with the results 
of the parameter studies, CS emission is not seen in the gas around 
\zoph, even when \hcop\ emission is relatively strong and somewhat 
weaker HCN emission is probably detected at a level some three times
weaker then \hcop.  However, CS emission is detected in the directions 
of several early-type stars used for optical
absorption line studies, with brightnesses that are quite comparable
to those of \hcop\ at similar \WCO\ in other directions.  This implies
somewhat larger variations in the CS/\hcop\ abundance ratio than were
seen in absorption, and larger values of X(CS).

To bridge the gap between these observations and the simulations
and parameter studies we calculated the brightnesses of \hcop\ and 
CS lines in diffuse cloud models that previously were used to calculate 
the brightness of CO.  In these models CO forms from the recombination 
of a fixed relative abundance X(\hcop) = $2 \times 10^{-9}$.  
This chemical relationship couples the 
emergent brightnesses of the two species and largely fixes the ratio 
of their line brightnesses that otherwise is difficult to adjust.  
The models reproduce the observed \hcop/CO brightness temperature ratio 
(1-2\%).

Plotting the observed brightness of polar molecules against \WCO\
creates degeneracies in the parametric dependences of the brightness
on other physical variables, such that curves of W(CS) vs. \WCO\ nearly 
coincide even for cloud models that differ appreciably in their 
hydrogen number and column densities (Fig. 4).  That is, for a fixed 
set of relative abundances,  the relative brightnesses are fixed over a wide 
range of physical parameter space.  In that case, the relative brightnesses
are determined chiefly by the various X(CS), X(HCN), etc, an insight
we also gained from the preceding LVG-based parameter studies. 

%Surveys of the galactic plane in species beside
%CO show that strongly-polar molecules (\hcop, HCN, CS, \cch) all appear
%with nearly equal brightness in CO clouds observed in the galactic 
%plane, and their brightnesses are 1-2\% that of CO  
%\citep{Lis95,HelBli97,McQSim+02}, similar to what is seen for \hcop\
%in the diffuse regime, but much larger than for CS or HCN.

\begin{acknowledgements}

The National Radio Astronomy Observatory is operated by Associated 
Universites, Inc. under a cooperative agreement with the US National 
Science Foundation.  I thank Alexandre Faure and Jonathan Tennyson 
for providing the full set of electron excitation rates for \hcop\ 
prior to publication and am happy to acknowledge valuable discussions
with Michel Guelin and Jerome Pety.  

This work was supported in part by the grant ANR-09-BLAN-0231-01 from 
the Agence Nationale de la Recherche (France), the SCHISM project
see http://schism.ens.fr/ .

\end{acknowledgements}
 
\bibliographystyle{apj}

\begin{appendix}

\section{Relative abudances of \hcop, HCN and CS}

Shown in Fig. A.1 are the column densities of \hcop, HCN and CS as observed
in absorption toward extragalactic mm-wave continuum sources in the work of
\cite{LucLis96} for \hcop, \cite{LisLuc01} for HCN and \cite{LucLis02} for CS.
%1
\begin{figure}
\includegraphics[height=6.2cm]{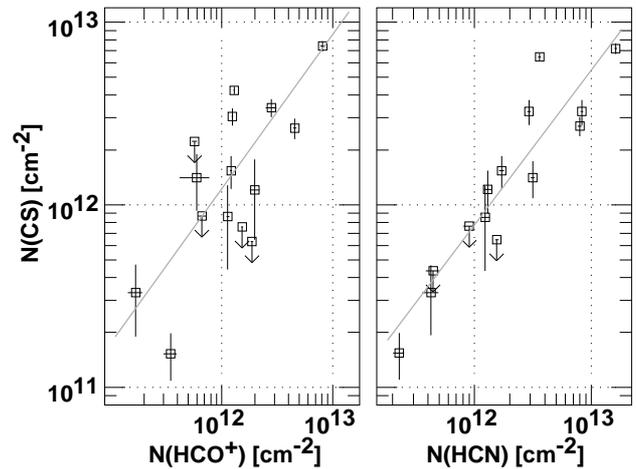}
\caption[]{CS column density of N(CS) plotted against N(\hcop) at left and
N(HCN) at right as observed in absorption toward extragalactic mm-wave
background sources, see Appendix A1.}
\end{figure}

\end{appendix}

\end{document}